\documentclass[11pt,a4]{article}
\usepackage{graphics}
\usepackage{comment}
\usepackage[numbers]{natbib}
\usepackage{amsmath}
\usepackage[a4paper,top=1.0cm,bottom=1.2cm,left=2cm,right=2cm,marginparwidth=1.5cm]{geometry}

\title { \bf Three Perspectives on Complexity -- \\ Entropy, Compression, Subsymmetry}
\author{Nithin Nagaraj$^\dagger$ and Karthi Balasubramanian$^*$}

\date{\small $^\dagger$Consciousness Studies Programme, National Institute of Advanced Studies\\ Indian Institute of Science Campus, Bengaluru 560 012, India. \\{\bf \small Email:~nithin@nias.iisc.ernet.in} \\
$^*$Department of Electronics \& Communication Engg., Amrita School of Engg., Coimbatore \\ Amrita Vishwa Vidyapeetham, Amrita University, India. \\{\bf \small Email:~b\_karthi@cb.amrita.edu}\\ October 30, 2016}

\begin{document}
\maketitle
\abstract{
There is no single universally accepted definition of `{\it Complexity}'. There are several perspectives on complexity and what constitutes complex behaviour or complex systems, as opposed 
to regular, predictable behaviour and simple systems. In this paper, we explore the following perspectives on complexity: {\it effort-to-describe} (Shannon entropy $H$, Lempel-Ziv complexity $LZ$), {\it effort-to-compress} ($ETC$ complexity) and {\it degree-of-order} (Subsymmetry or $SubSym$). While
Shannon entropy and $LZ$ are very popular and widely used, $ETC$ is a recently proposed measure for time series. In this paper, we also propose a novel normalized measure $SubSym$ based on the existing idea of counting the number of subsymmetries or palindromes within a sequence. We compare the performance of these complexity measures on the following tasks: a) characterizing complexity of short binary sequences of lengths 4 to 16, b) distinguishing periodic and chaotic time series from 1D logistic map and 2D H\'{e}non map, and c) distinguishing between tonic and irregular spiking patterns generated from the `Adaptive exponential integrate-and-fire' neuron model. Our study reveals that each perspective has its own advantages and uniqueness while also having an overlap with each other. 
%
%
} 
%
\section{Introduction}
\label{intro}
The seemingly naive procedure of finding decimal expansion of a real number, synchronized beating of heart cells, irregular firing of neurons in the brain, global economy and society, motion of planets and galaxies and the everyday weather - are but a few examples of complex systems.  While there is little agreement on the precise mathematical definition of {\it complexity}, there is no doubt on the ubiquity of complex systems in nature and society.

The ability to model, analyze and understand complex systems is invaluable since it allows us to control, predict and exploit complex phenomena to our advantage. For example, by characterizing the complexity of the chaotic nature of cardiovascular dynamics, we may one day understand the effects of aging~\cite{CardiacAging1992}. Understanding complexity may be the key to unlocking the mysteries of the brain since chaotic behaviour is abound in the central nervous system~\cite{BrainChaos2001,BrainChaos2001b}. Complex behaviour of neural signals may be even necessary for multiplexing a large number of neural signals in the brain for successful communication in the presence of neural noise and interference~\cite{NagarajCSMultiplex}. Pseudo-random sequences having high linear (and nonlinear) complexity, but easy to generate in hardware, are employed in various cryptographic protocols~\cite{pseudorandomness}.  Complexity and randomness is increasingly playing an important role in psychology to decipher behaviours and mental processes pertaining to memory (see~\cite{Pscyhology1} and references therein),  cognitive effects of aging and human judgments of visual complexity~\cite{Toussaint}. Such a list is by no means exhaustive, but only serves to indicate the widespread prevalence and use of the notion of complexity.

There are broadly two categories of complex systems - deterministic and non-deterministic~\footnote{Most naturally occurring systems are a hybrid since stochastic noise is inevitable.}. Chaotic systems are examples of deterministic complex systems which are capable of producing rich and complex behaviour ranging from periodic to quasi-periodic to completely random-looking and unpredictable outputs. Stochastic systems are inherently non-deterministic (`random') and capable of exhibiting complex dynamics which can be predicted only probabilistically.  In either case, measuring complexity (of time series measurements) is an important step in characterizing or modeling the system. In this paper, we concern ourselves with the all-important question: {\it What is complexity?}

\subsection*{Complexity: many facets}
There is no single universally accepted definition of `{\it Complexity}'. There are several perspectives on complexity and what constitutes complex behaviour or complex systems, as opposed 
to regular, predictable behaviour and simple systems. With the recent availability of high speed and low cost computational tools at our disposal, analysis of complex systems has become data-intensive. Lloyd~\cite{lloyd2001} lists three questions which researchers ask in order to quantify the complexity of the thing (object) under study: 

\begin{enumerate}
\item How hard is it to describe? ({\it effort-to-describe})
\item How hard is it to create? ({\it effort-to-create})
\item What is its degree of organization? ({\it degree-of-order}).  
\end{enumerate}

We have indicated these questions with the corresponding italicized phrases in parenthesis (not present in Lloyd's original formulation). In this paper, we study three perspectives of complexity namely: {\it effort-to-describe}, {\it effort-to-compress} and {\it degree-of-order}. We have chosen a new perspective `{\it effort-to-compress}' instead of `{\it effort-to-create}' and our interpretation of {\it degree-of-order} is different from Lloyd's (as will be evident in this paper). 

By no means, is this exhaustive and there are several other perspectives which escape us for the moment. Table~\ref{tab:table3Perspectives} gives the definitions, examples of these three perspectives and their contrasting approaches to characterize {\it complexity} of an object. In the rest of the paper, our goal is to compare and evaluate specific measures that are representative of these three perspectives. 

\begin{table}[!ht]
\centering
\caption[ThreePerspectives]{Three perspectives on `{\it Complexity}'.} %
\label{tab:table3Perspectives} %
\vspace{0.1in}
\begin{tabular}{|p{2.0cm}|p{3cm}|p{2.8cm}|p{5cm}|}  
\hline
{\bf Perspective} & {\bf Definition} & {\bf Examples} & {\bf Remarks} \\
\hline
{\it Effort-to-describe} & The number of bits needed to describe the object, the length of the shortest computer program to describe the object, the size of the dictionary that captures the minimum exhaustive history of the object. & Shannon Entropy ($H$), Kolmogorov-Chaitin complexity, Lempel-Ziv complexity ($LZ$) & Shannon entropy is measured in bits. $LZ$ is popularly used in place of Kolmogorov-Chaitin complexity which is difficult to compute. $LZ$ is related to the universal lossless compression algorithm by the same name, $LZ$ complexity is used extensively in several applications.\\
\hline
{\it Effort-to-compress} & The effort to losslessly compress the object by a pre-determined lossless compression algorithm.  & Effort-To-Compress ($ETC$), a recently proposed measure, is based on the lossless compression algorithm known as Non-Sequential Recursive Pair Substitution (NSRPS). &  $ETC$ could be defined with other lossless compression algorithms as well. The {\it effort-to-describe} of two objects may be the same, and yet admit different {effort-to-compress} complexity and vice versa.\\
\hline
{\it Degree-of-order} & The amount of {\it order} or {\it structure} that is {\it intrinsic} to the object. Symmetry is a good candidate to capture this notion. & Subsymmetries ($SubSym$) & $SubSym$ as a normalized complexity measure (between 0 and 1) for time series is defined for the first time in this paper. $SubSym$ is based on Alexander and Carey's idea of counting number of subsymmetries (palindromes) as a measure of visual complexity of binary strings.\\
\hline
\end{tabular}
\end{table}
\section{The Three Perspectives}
In this section, we give in-depth definitions and descriptions of the four measures that together make up the three perspectives. Among the four measures, two of the measures are well known and widely used in several applications (Shannon entropy and Lempel-Ziv complexity). The third complexity measure known as Effort-To-Compress is a recently introduced measure to characterize complexity of time series and is beginning to be used in several applications. The fourth measure (subsymmetries) is a new measure that we propose in this paper for characterizing complexity of time series, though the underlying idea of counting subsymmetries (palindromes) is not new.   

\subsection{Effort-to-describe: $H$ and $LZ$}
Claude Shannon's notion of {\it entropy} as a quantitative measure of information content~\cite{Shannon1948} is very popular and widely used to characterize the complexity of a given time series. The origins of the idea can be traced back to thermodynamics (Clausius, 1965) and in statistical physics (Boltzmann and Gibbs, 1900s). Shannon entropy\footnote{We shall always refer to first-order empirically estimated Shannon entropy throughout the paper. Here `first-order' refers to the fact that the entropy is computed on the first-order probability distribution, treating symbols as independent of each other.} of a discrete random variable $\chi$ is defined as:
\begin{equation}
    H(\chi) = - \sum_{i=1}^{M} p_i \log_2(p_i)~~~\mbox{bits/symbol}, \label{eq:eqnENT}
\end{equation}
where $\chi$  takes $M$ possible events with the probability of occurrence of the $i$-th event given by $p_i>0$. The maximum value of the concave function $H(prob.)$ is achieved for a uniform random variable with all events equally likely and is equal to $H_{max} = \log_2(M)$ bits.
 
One simple yet illuminating interpretation of Shannon entropy is by means of Yes/No questions~\cite{CoverThomasBook}. Entropy can be defined as the {\it minimum} expected number of Yes/No questions needed to remove the uncertainty in the random experiment. Thus, entropy serves as measure of the {\it effort-to-describe} the outcome of a random experiment. 

Shannon entropy and other information theoretic measures are widely used in communication, data compression, error control coding, cryptography, biomedical applications, financial time series analysis, non-linear dynamics and as various entropic forms in physics (please refer to \cite{BookChapter} and references there in).


\subsection*{Kolmogorov-Chaitin Complexity and Lempel-Ziv Complexity}
Another very important measure of complexity goes by the name Kolmogorov complexity (1963) which is defined as the length of the shortest computer program (in a pre-determined chosen programming language) that produces the given string/sequence as output. It is also known as {\it descriptive complexity}, {\it Kolmogorov-Chaitin complexity}, {\it algorithmic entropy},  or {\it program-size complexity}. Please refer to Li and Vitanyi~\cite{LiVitanyi2008} for further details.

Owing to the uncomputability of Kolmogorov complexity, lossless compression algorithms have been used to approximate as an upper bound. Lempel-Ziv complexity is one such measure~\cite{LZComplexity} that is closely related to Lempel-Ziv universal lossless compression algorithm~\cite{LZCompression}. It is defined as the size of the dictionary which captures the minimum exhaustive history of the input sequence, from which the entire sequence can be uniquely expressed (or compressed). For example, if the input string is `$aacgacga$', the dictionary is built up of words that form the minimum exhaustive history as the input sequence is parsed from left to right. In this example, the dictionary is $\{ a, ac, g, acga \}$. Thus the value of Lempel-Ziv complexity ($LZ$) is the size of the dictionary which is $c(n) = 4$. In order to allow us to compare $LZ$ complexities of strings of different length, a normalized measure is proposed~\cite{Aboy2006} as $C_{LZ}  = (c(n)/n)\log_\alpha (n)$, where $n$ is the length of the input sequence, $\alpha$ is the number of unique symbols in the input sequence. 

Lempel-Ziv complexity (LZ) is widely used in a variety of applications such as estimation of entropy of spike trains, biomedical applications, financial time series analysis, bioinformatics applications,  analysis of cardiovascular dynamics and several others (please refer to \cite{BookChapter} and references there in).

\subsection{Effort-to-compress: $ETC$}
Recently~\cite{ETCpaper}, we have proposed a new complexity measure which has a different perspective on complexity. Known  as the {\it Effort-To-Compress} or ETC, it measures the effort needed to compress the input sequence by a lossless compression algorithm. Specifically, we employ the Non-Sequential Recursive Pair Substitution  (NSRPS) algorithm~\cite{NSRPSpaper}. The algorithm is simple to describe. ETC proceeds by identifying that pair of consecutive symbols in the input sequence that is most frequently occurring  and replaces it with a new symbol. In the next pass, this is repeated. In this fashion, at every iteration the length of the sequence reduces and the algorithm stops when the sequence turns into a constant sequence (this is guaranteed as the length decreases in successive iteration). The ETC complexity measure is defined as the number of iterations needed to reduce the input sequence into a constant sequence. As an example, ETC transforms the input sequence $aacgacga \mapsto a1g1ga \mapsto 2g1ga \mapsto 31ga \mapsto 4ga \mapsto 5a \mapsto 6$. The value of the complexity measure in this instance is $N=6$. The normalized value of the measure is given by $\frac{N}{L-1}$ where $L$ is the length of the input sequence (0.8571 for this example). Note that $0 \leq \frac{N}{L-1} \leq 1$.  Lower the value of the measure, lower is the complexity. 
 
ETC was originally proposed for automatic identification and classification of short noisy sequences from chaotic dynamical systems~\cite{ETCpaper}. It has recently been applied for characterizing dynamical complexity of time series from chaotic flows and maps, as well as two-state markov chains~\cite{BookChapter, MarkovEPJST}, analyzing short RR tachograms from healthy young and old subjects~\cite{PeerJ}, and for developing a measure of integrated information~\cite{PLOSONE}. 
 
\begin{table}[!h]
\centering
\caption[SS]{$SubSym$: A normalized complexity measure defined using number of subsymmetries (palindromes)~\cite{AlexanderCarey}. As an example we consider three binary strings of length 6. The normalized measure is defined with respect to the all-zero sequence $Z$ for which the value of the measure is 0.} %
\label{tab:tableSubSym} %
\vspace{0.1in}
\begin{tabular}{|c|c|c|c|}
  \hline
  String & Length of substring & Number of palindromes & $SubSym$\\
   \hline
           & 2 & 5  &   \\
    $Z=$       & 3 & 4  &   \\
  `000000' & 4 & 3  &   \\
           & 5 & 2  &   \\
           & 6 & 1  &   \\
           &  & Total$=15$  &  $1 - \frac{15}{15} = 0$  \\
\hline
\hline
           & 2 & 1  &   \\
           & 3 & 2  &   \\
  `010010' & 4 & 1  &   \\
           & 5 & 0  &   \\
           & 6 & 1  &   \\
           &  & Total$=5$  &  $1 - \frac{5}{15} = 0.6667$  \\
\hline
\hline
           & 2 & 1  &   \\
           & 3 & 2  &   \\
  `011010' & 4 & 1  &   \\
           & 5 & 0  &   \\
           & 6 & 0  &   \\
           &  & Total$=4$  &  $1 - \frac{4}{15} = 0.7333$  \\
    \hline
\end{tabular}
\end{table}
 
\subsection{Degree-of-order: $SubSym$}
In the original taxonomy of complexities by Lloyd~\cite{lloyd2001}, degree of organization was divided into two aspects - the difficulty of describing organizational structure, and the amount of information that is shared between the parts of a system as a result of this organizational structure. We interpret this perspective in a slightly different way. We are interested in the {\it degree-of-order} that is {\it intrinsic} to an input sequence. The best candidate for such a notion is {\it symmetry}.

The idea of counting the total number of subsymmetries that is intrinsic in a binary string goes back to Alexander and Carey~\cite{AlexanderCarey} who were interested in characterizing the cognitive simplicity of binary patterns. However, it has largely been forgotten, only recently revived by Toussaint {\it et al.}~\cite{Toussaint}, again in the context of human judgment of visual complexity. The definition is very simple - just count the total number of all substrings of lengths $ \geq 2$ which are {\it palindromes}. 

With the already existing notion of subsymmetries (palindromes), we define a new normalized complexity measure for sequences as follows\footnote{The length of the sequence we consider is always $n>1$.}:
\begin{eqnarray*}
SubSym(x) & = &  1 - \frac{subsymmetries(x)}{subsymmetries(Z)}, \\
& = &  1 - \frac{subsymmetries(x)}{ \bigg (\frac{n(n-1)}{2} \bigg )},
\end{eqnarray*}
where $x$ is the input sequence, $Z$ is the all-zero sequence and the function {\it subsymmetries(.)} returns the total number of subsymmetries or palindromes. By a matter of simple counting, it is easy to see that an all-zero sequence of length $n$ has total number of subsymmetries or palindromes $= \frac{n(n-1)}{2}$. The idea behind this definition is that the all-zero sequence has the {\it highest order} in that it has the most number of subsymmetries. Hence, the number of subsymmetries of the input sequence $x$ is divided by that of the all-zero sequence. This gives the proportion of order or simplicity in the input sequence. In order to give a measure of complexity we subtract this quantity from 1. The resulting normalized measure satisfies $0 \leq SubSym(x) \leq 1$.

As an example, we consider three binary strings of length 6 (refer to Table~\ref{tab:tableSubSym}) and compute the normalized measure: $SubSym$. As per the measure, the string `$000000$' has the highest order (15 palindromes), followed by the string `$010010$' (5 palindromes) and `$011010$' (4 palindromes). The normalized complexity measure $SubSym$ for the three strings are $0$, $0.6667$ and $0.7333$ respectively.

\section{Comparing The Three Perspectives}

In this section, we rigorously evaluate the four complexity measures (just described) arising from the three perspectives. To this end, we consider the following three tasks:
\begin{itemize}
\item Characterizing complexity of short binary sequences.
\item Distinguishing periodic and chaotic time series.
\item Distinguishing tonic and irregular spiking from a spiking neuron model.
\end{itemize}

One important point to note is that all these measures work with symbols (such as `$0$' and `$1$', $\{a, c, g, t\}$, etc.). So, we need a procedure to convert the real-valued time series into a {\it symbolic sequence} (a sequence of symbols). This is done by first defining a {\it partition} and assigning unique symbols to non-overlapping bins which form the partition. For example, consider a real-valued time series that takes values in the range $[Min, Max]$. Let us define a partition consisting of 2 bins: $[Min, Mid)$ and $[Mid, Max]$ where $Mid = \frac{Min+Max}{2}$. Thus every value of the time series is checked to see in which bin it belongs. If it belonged to the bin (interval) $[Min, Mid)$, it is coded as `$0$' and if it belonged to $[Mid, Max]$ it is coded as `$1$'. Thus, the resulting symbolic sequence will contain symbols made up of $0$s and $1$s and is now ready for analysis using the complexity measures. 

\subsection{Complexity of short binary sequences}
Binary sequences or strings arise in several applications. For example, in analyzing membrane potential waveforms (also known as spike trains) which are obtained from neurons, the signal is first converted into a symbolic sequence of zeros and ones indicating `no-spiking' and `spiking' in a moving window of a pre-determined length~\cite{Amigo2004}. The resulting binary sequence is then analyzed in order to estimate its entropy by means of Lempel-Ziv complexity~\cite{Amigo2004}. In several cases, the neuron under study may fire only a few times ($10-20$ or even lesser) in the window chosen for the study. Thus, it is vital to consider the performance of complexity measures on short binary sequences.

\begin{table}[!h]
\centering
\caption[FourMeasures]{Four binary strings of length 4 and their respective complexities.} %
\label{tab:table4Measures} %
\vspace{0.1in}
\begin{tabular}{|c|c|c|c|c|}
\hline
String & $H$ (bits) & $LZ$ & $ETC$ & $SubSym$\\
\hline
`000000' & 0 & 0.8617 & 0 & 0\\
\hline
`010010' & 0.9183 & 1.723 & 0.4 & 0.6667\\
\hline
`011010' & 1.0 & 1.723 & 0.8 & 0.7333\\
\hline
`011100' & 1.0 & 1.723 & 1.0 & 0.6667\\
\hline
\end{tabular}
\end{table}

As an example, consider the four binary strings in Table~\ref{tab:table4Measures} and the corresponding values of the four complexity measures. Intuitively, we expect that binary strings with more {\it structure} or {\it order} have a lower value of complexity (eg. `$000000$' and `$010010$') than those with less structure (eg. `$011010$' and `$011100$'). From the table above, we can see that different measures capture different aspects of {\it order} and {\it structure}. Shannon entropy ($H$) is blind to the arrangement of 0s and 1s and is only sensitive to their relative counts. $LZ$ performs very poorly as it has issues with short data-lengths which has been previously noted~\cite{Hu2006}. $ETC$ and $SubSym$ seem to yield reasonable values for this example. $SubSym$ puts `$011010$' as the string with highest complexity since it has the least number of palindromes/subsymmetries. The number of subsymmetries in `$010010$' and `$011100$' are the same and hence have identical values of $SubSym$. For $ETC$, `$011100$' has slightly higher complexity than `$011010$' since the latter has two patterns of same frequency (`$01$' and `$10$') whereas the former has only one pattern with  highest frequency (`$11$').

The first test that we perform is to characterize the complexity of all possible binary strings (or sequences) of lengths ranging from 4 to 16. One of the metrics that we use to evaluate these measures is the number of distinct complexity values the complexity measure takes across the entire ensemble of binary strings of a particular length.  A larger number of distinct values is desirable since it allows to distinguish complexities of individual binary strings more finely. 

\begin{figure}[!h]
\centering
\resizebox{1.1\columnwidth}{!}
{
\includegraphics{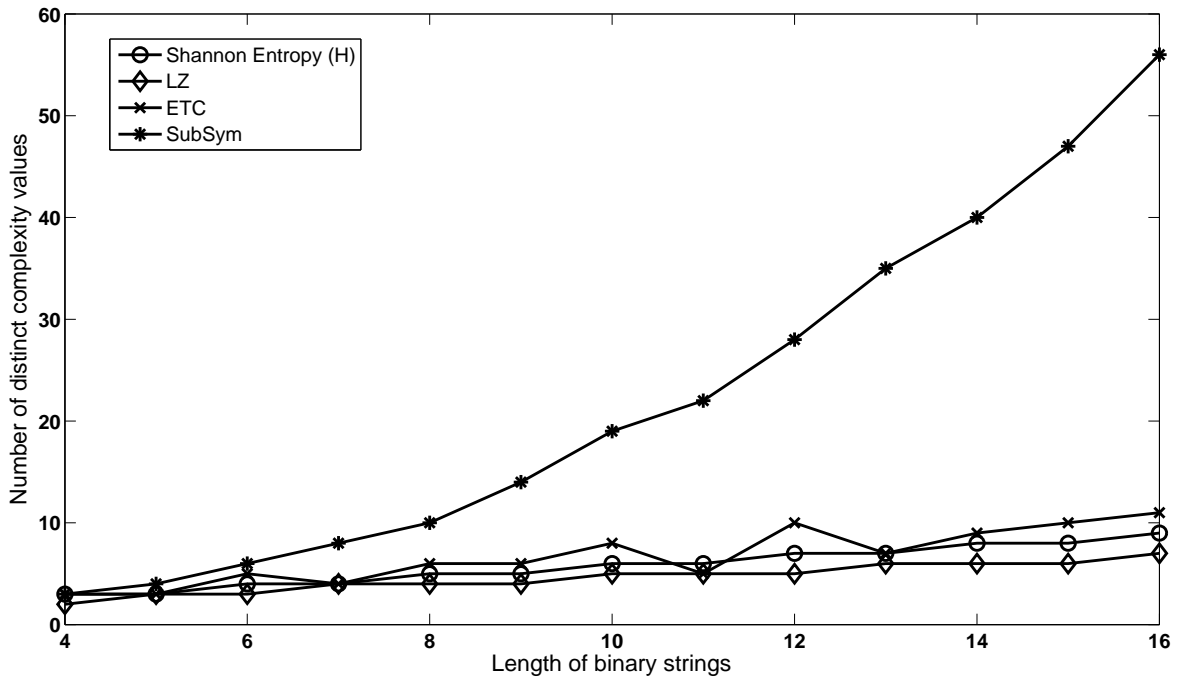}
}
\caption{The number of distinct complexity values for all binary strings of lengths 4 to 16, for each of the four measures ($H$, $LZ$, $ETC$ and $SubSym$). As it can be seen, $SubSym$ has the most number of distinct values followed by $ETC$, $H$ and $LZ$. A larger number of distinct values is desirable since it allows to distinguish complexities of different individual binary strings more finely.}
\label{figure:Binseq_res1}
\end{figure}

Fig.~\ref{figure:Binseq_res1} shows the graph of the number of distinct complexity values (for the four measures) for all binary strings of lengths 4 to 16. As it can be seen, $SubSym$ has the most number of distinct values followed by $ETC$, $H$ and $LZ$. It appears that among the measures considered, $SubSym$ is ideally suited for characterizing the complexity of short binary strings. However, this alone is insufficient, and the next test will determine whether these measures can successfully distinguish between periodic and chaotic time series.

\subsection{Distinguishing periodic and chaotic time series}
In several applications, there is a need to distinguish between periodic behaviour and irregular or chaotic behaviour. For example, it is well known that 
heart rate in healthy human hearts 	is not constant or periodic, but shows irregular patterns that are typically associated with a chaotic dynamical system~\cite{Goldberger1991}. Several studies suggest that aging and other pathological physiological conditions result in reduced complexity of heart dynamics, leading to a more regular heart rate variability~\cite{CardiacAging1992, PathologyHeart}. 

We are interested in determining whether the four complexity measures ($H$, $LZ$, $ETC$ and $SubSym$) are able to distinguish between chaotic (irregular) and periodic time series. In order to test this, we first consider the 1D chaotic logistic map~\cite{Alligood} given by the following equation: 
\begin{eqnarray*}
\text{Logistic map:~}    x_{n+1} & = & a x_{n}(1-x_{n}),
\end{eqnarray*}
where $n$ is the discrete time sample, $\{ x_i \}_{i=1}^{i=LEN}$ is the time series of length $LEN$ and $a$ is the bifurcation parameter. The values of $a$ are chosen such that the resulting time series is periodic ($a=3.83$), weakly chaotic ($a=3.75$), strongly chaotic ($a=3.9$) and fully chaotic\footnote{Fully chaotic refers to the absence of attracting periodic orbits and the Lyapunov exponent is positive and reaches maximum at $a=4.0$ indicating the highest degree of chaos.} ($a=4.0$). The length of the time series generated is varied as $LEN = 20, 40, 80, 100, 150$ and $200$.  For each of these lengths, 50 different sets of time series are generated from distinct initial conditions chosen randomly from the interval $(0, 1)$. The resulting ensemble of time series is converted into a symbolic sequence with two symbols: `$0$' corresponding to the interval $(0, 0.5]$ and `$1$' corresponding to interval $[0.5, 1)$. These symbolic sequences are subsequently analyzed by the four complexity measures. 
\begin{figure}[!h]
\centering
\resizebox{1.1\columnwidth}{!}
{
\includegraphics{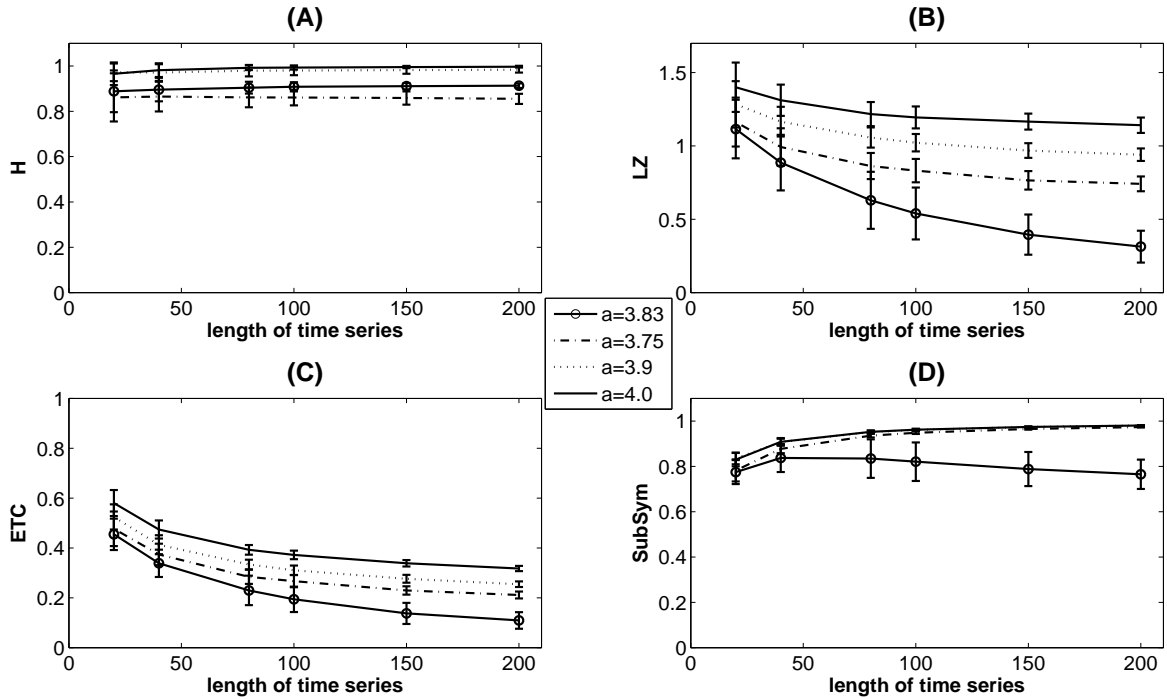}
}
\caption{Mean complexity measures vs. length of time series for the logistic map ($a=3.75$, $3.83$, $3.9$ and $4.0$) over 50 randomly chosen initial conditions (and 2 bins) for (A) Shannon entropy $H$, (B) Lempel-Ziv complexity $LZ$, (C) Effort-To-Compress complexity $ETC$, and (D) $SubSym$ complexity. Errorbars indicate one standard deviation. $LZ$ and $ETC$ show very good separation for the four time series as their length increases.}
\label{figure:4measuresChaos}
\end{figure}
\par Fig.~\ref{figure:4measuresChaos} shows the mean complexity values vs. length of time series. We make the following observations. Mean Shannon entropy values are fairly constant as the length of the time series is varied. Though entropy is able to separate between the chaotic and periodic time series, it does poorly to distinguish between strong chaos ($a=3.9$) and full chaos ($a = 4.0$). Both mean $LZ$ and mean $ETC$ values show good separation for the four cases, especially with increasing lengths.  $SubSym$ is able to clearly separate the periodic ($a=3.83$) from the chaotic, though it is unable to distinguish between weak chaos ($a=3.75$), strong chaos ($3.9$) and full chaos ($a=4.0$).

\par For the next experiment, we are interested in determining the minimum length of time series needed by these measures to classify periodic time series from chaotic time series. We consider two non-linear chaotic maps - the 1D logistic map (defined previously) and the 2D H\'{e}non map~\cite{Alligood} given by:
\begin{eqnarray*}
\text{H\'{e}non map:~}   x_{n+1} & = & 1-ax_n^2+y_n, \text{~~~} y_{n+1}  =  bx_n.
\end{eqnarray*}
For this experiment, we consider only two choices for the parameter $a$ for the 1D logistic map: $a = 3.83$ and $a = 4.0$ which correspond to periodic and chaotic behaviour respectively. Similarly, for the 2D H\'{e}non map, with the parameter $b$ fixed at 0.3, the two choices of parameter $a$ are $a = 1.3$ and $a = 1.4$ corresponding to periodic and chaotic behaviour respectively. As before, we take 2 equal sized bins for creating the symbolic sequence, and 50 trials. The aim of this exercise is to determine that length of the time series, below which the measure fails to distinguish between periodic and chaotic cases. 

\begin{table}[!h]
\centering
\caption[tableChaos]{Minimum length of time series required for statistically distinguishing between periodic and chaotic time series (obtained from randomly chosen initial conditions over 50 trials) from the chosen chaotic maps (logistic and H\'{e}non). $ETC$ requires the least length of time series.} %
\label{tab:tableMINLEN} %
\vspace{0.1in}
\begin{tabular}{|c|c|c|c|c|}
\hline
Chaotic & Min. Len. & Min. Len. & Min. Len. & Min. Len.\\
Map & $H$ & $LZ$ & $ETC$ & $SubSym$\\
\hline
Logistic Map & $15$ & $8$ & $6$ & $10$\\
\hline
H\'{e}non Map& $>2000$ & $30$ & $15$ & $>2000$\\
\hline
\end{tabular}
\end{table}

Table~\ref{tab:tableMINLEN} shows the minimum length of time series required to statistically distinguish between periodic and chaotic time series for the four measures. We can infer that - for the 1D logistic map, all the measures succeed in discriminating periodic from chaotic time series, though they need different minimum lengths to do the job. For the 2D H\'{e}non map, both Shannon entropy and $SubSym$ fail to discriminate even for lengths as large as 2000. $ETC$ has the best (lowest {\it minimum length}) performance for both maps.

We now consider the final experiment of distinguishing between tonic (regular) spiking patterns and irregular (chaotic) spiking patterns from a spiking neuron model.

\subsection{Spiking neuron model}
There are $\approx86$ billion neurons in the human central nervous system. These cells are the basic information processing units and they form a highly complex network. These neurons respond to a stimulus (either an external or an internal stimulus) by firing different patterns of short duration cell membrane potential spikes (known as `action potentials'). These carry important information about the stimuli~\cite{Gutig2014}.

In order to investigate the dynamics of neuronal behaviour, different models of their firing have been proposed. One such single spiking neuron model is the `Adaptive Exponential Integrate-and-Fire' (AdEx) model, initially proposed by Brette and Gerstner~\cite{AdEx2005}. This is a simple and realistic model used commonly and has only two equations and a reset condition. The different firing patterns produced by this model are described in Naud {\it et al.}~\cite{Naud2008}. The equations governing the model are:
\begin{eqnarray*}
C \frac{dV}{dt} & = & -g_L (V - E_L) + g_L \Delta_T \exp \Bigg ( \frac{V-V_T}{\Delta_T} \Bigg ) + I -w,\\
\tau_w \frac{dw}{dt} & = & a(V - E_L) - w.
\end{eqnarray*}

\begin{figure}[!h]
\centering
\resizebox{0.9\columnwidth}{!}
{
\includegraphics{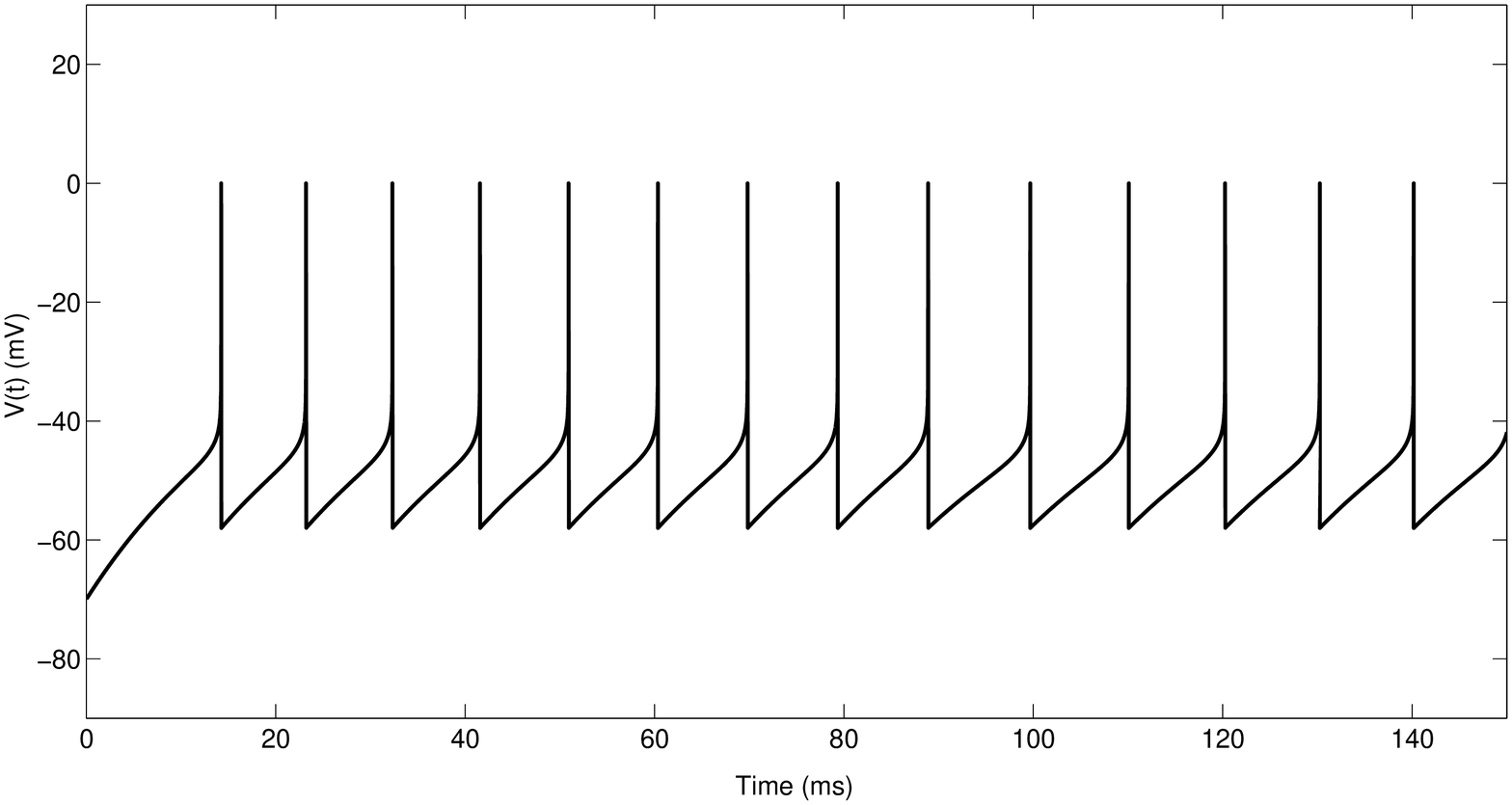}
}
\resizebox{0.9\columnwidth}{!}
{
\includegraphics{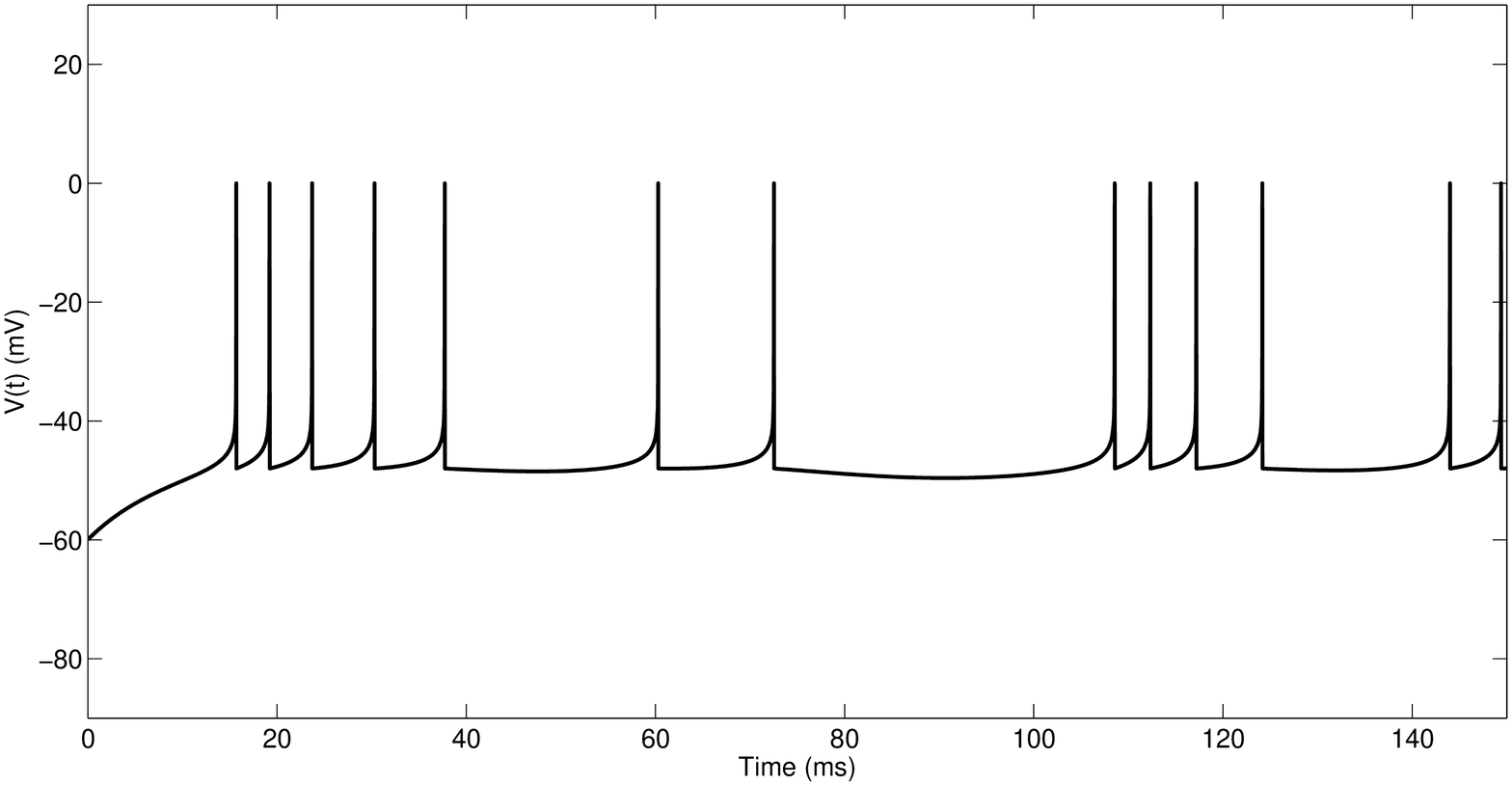}
}
\caption{Membrane potential $V(t)$ (mV) vs. time (ms) from the `Adaptive Exponential Integrate-and-Fire' (AdEx) neuron firing model used for the study. Top: Tonic spiking waveform showing periodic spikes. Bottom: Irregular spiking waveform showing chaotic behaviour.}
\label{figure:AdExTonicIrr}
\end{figure}

\par In the above equations, $V(t)$ is the membrane potential, $I(t)$ is the injection current and $w$ is the adaptation current. The model has 9 parameters: total capacitance ($C$), total leak conductance ($g_L$), effective rest potential ($E_L$), threshold slope factor ($\Delta_T$), effective threshold potential ($V_T$), conductance ($a$), time constant ($\tau_w$), spike triggered adaptation ($b$) and reset potential ($V_r$). When the current drives the voltage beyond the threshold value of $V_T$, the exponential term causes a sharp increase (upswing) in the action potential. The rise is stopped at reset threshold that has been fixed at 0 mV. The decrease (downswing) in the action potential is represented by the reset condition: $\text{if } V > 0 \text{ mV then } V \rightarrow V_r, w \rightarrow w_r \text{ where } w_r = w+b.$ 

\begin{figure}[!h]
\centering
\resizebox{1.1\columnwidth}{!}
{
\includegraphics{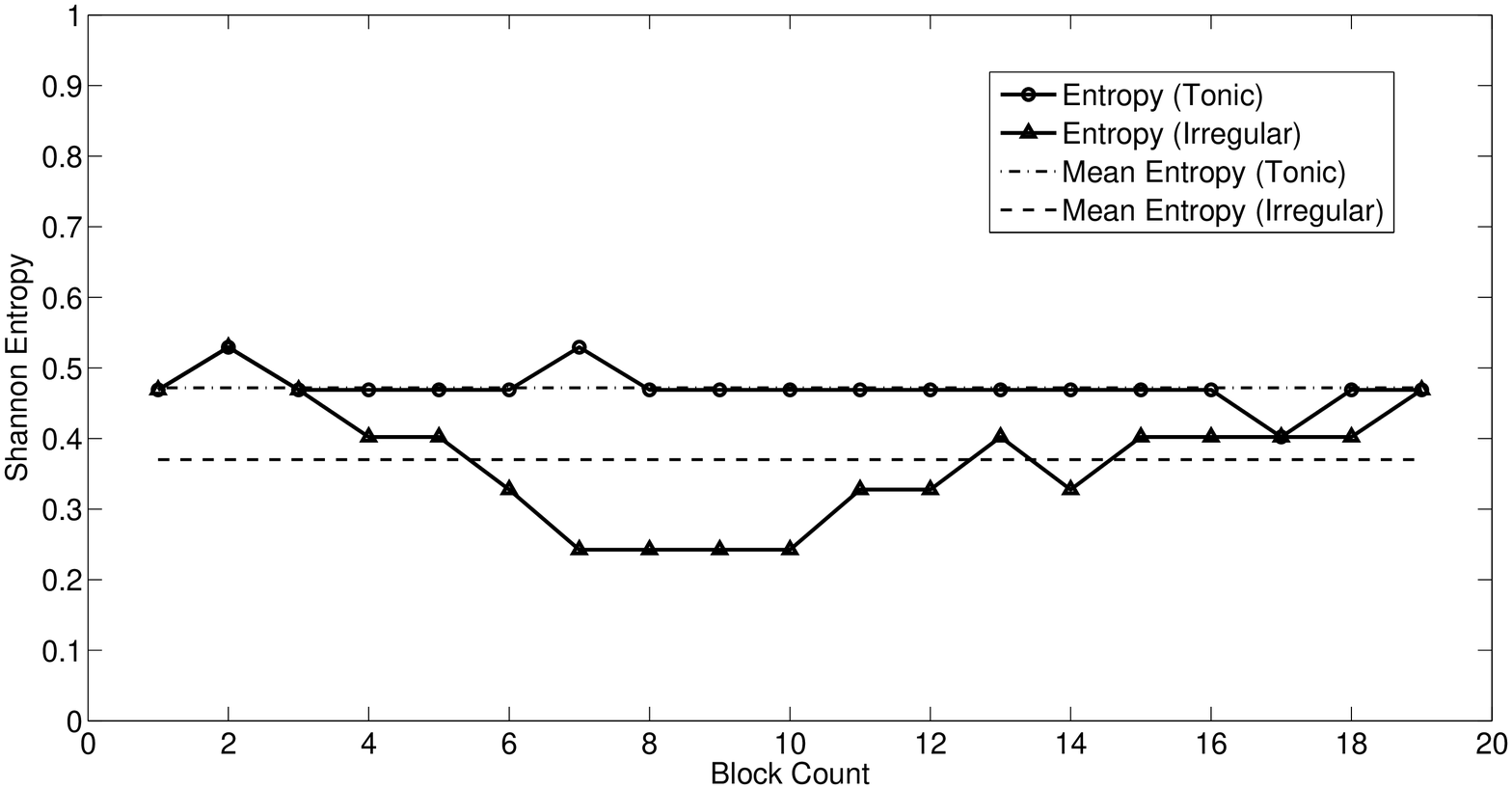}
}
\caption{Shannon entropy of tonic and irregular spiking using a moving window of 50 ms with 90\% overlap. Mean entropy is higher for tonic than that of irregular spiking which is not intuitive.}
\label{figure:AdExEntropy}
\end{figure}
\begin{figure}[!h]
\centering
\resizebox{1.1\columnwidth}{!}
{
\includegraphics{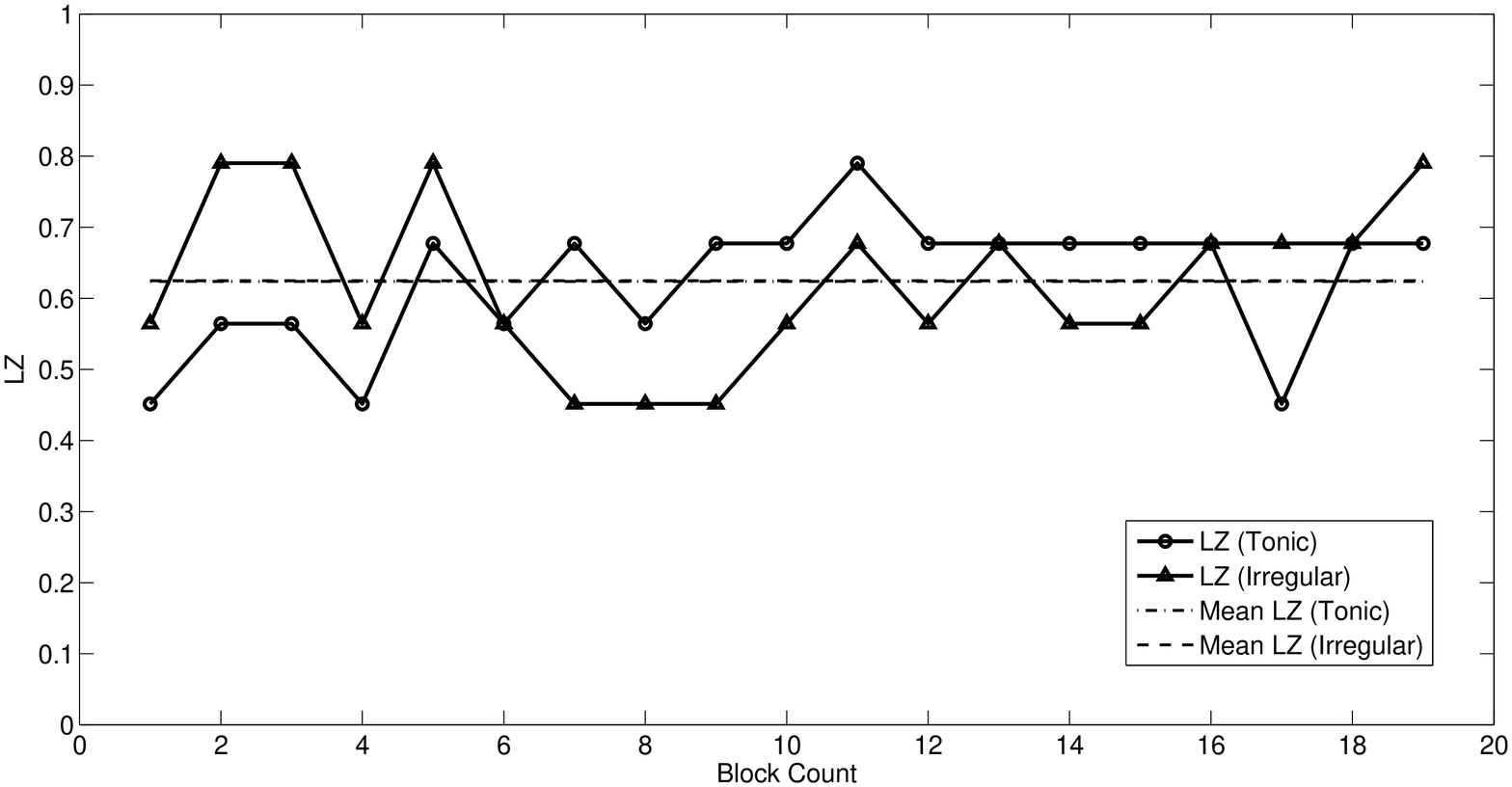}
}
\caption{LZ complexity measures of tonic and irregular spiking using a moving window of 50 ms with 90\% overlap. Mean LZ for tonic and irregular spiking are similar indicating the inability of LZ complexity measure to distinguish between the two spike trains.}
\label{figure:AdExLZ}
\end{figure}
\begin{figure}[!h]
\centering
\resizebox{1.1\columnwidth}{!}
{
\includegraphics{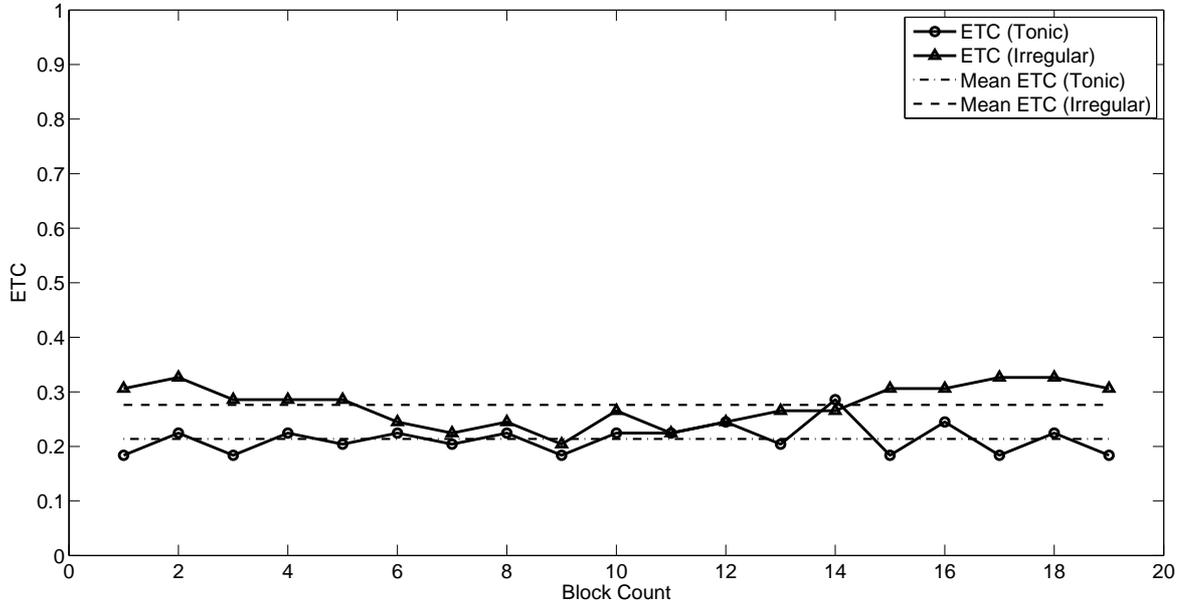}
}
\caption{ETC complexity measures of tonic and irregular spiking using a moving window of 50 ms with 90\% overlap. There is a clear difference for the two spiking patterns. The mean ETC for irregular spiking is greater than mean ETC for tonic spiking indicating that ETC is able to distinguish between the two spike trains.}
\label{figure:AdExETC}
\end{figure}
\begin{figure}[!h]
\centering
\resizebox{1.1\columnwidth}{!}
{
\includegraphics{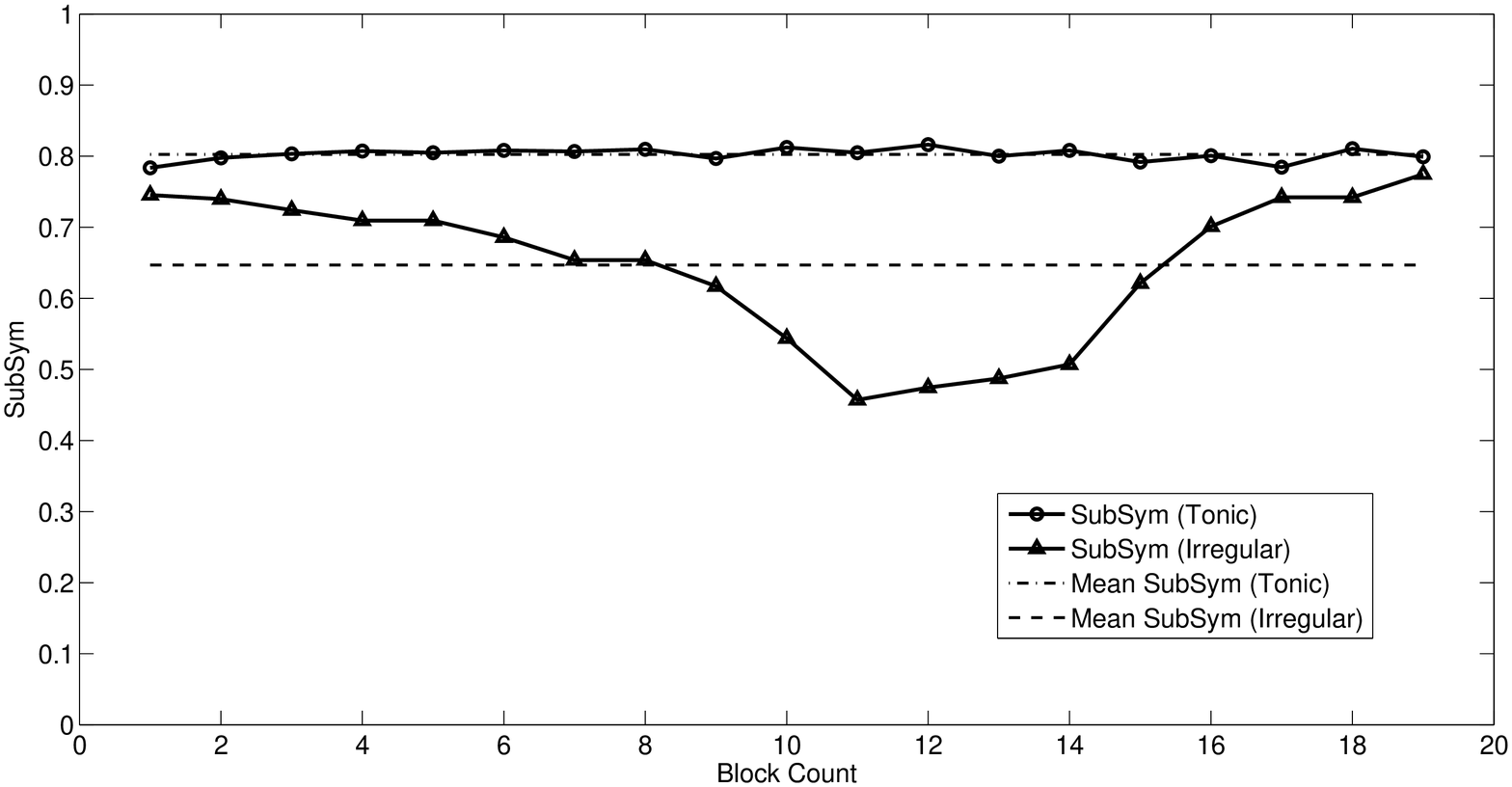}
}
\caption{$SubSym$ complexity measure of tonic and irregular spiking using a moving window of 50 ms with 90\% overlap. Mean $SubSym$ is higher for tonic than that of irregular spiking which is not intuitive.}
\label{figure:AdExSubSym}
\end{figure}
By varying the 9 parameters, various firing patterns can be obtained~\cite{Naud2008}, such as - tonic spiking, adaptation, initial burst, regular bursting, delayed accelerating, delayed regular bursting, transient spiking and irregular spiking. We are interested in two firing patterns namely: {\it tonic spiking} and {\it irregular spiking}. These are shown in Fig.~\ref{figure:AdExTonicIrr}. In {\it tonic spiking}, spikes are produced at regular intervals, whereas the interspike intervals in {\it irregular spiking} are aperiodic and is also highly sensitive to initial conditions (hallmark of chaos). Thus discriminating between these two patterns is akin to distinguishing periodic signals from chaotic ones. 

We follow the procedure of \cite{Amigo2004} for converting our waveforms into symbolic sequences of $0$s and $1$s. The entire waveform is partitioned into non-overlapping time windows of length $1$ ms. In each window, if one or more spikes are present, it is coded as `$0$', else coded as `$1$'. The four complexity measures ($H$, $LZ$, $ETC$ and $SubSym$) are estimated for the symbolic sequence in a moving time window of 50 ms (with an overlap of $90\%$ between successive windows) and the results are shown in Fig.~\ref{figure:AdExEntropy}-\ref{figure:AdExSubSym}.  

We make the following observations based on our results:

Across both the spike trains, it can be observed that LZ shows similar variations while ETC shows greater variation for the irregular spike train as compared to the tonic case.  Further, statistical evaluation shows that the mean LZ values of both tonic and irregular spike trains are same while there is an appreciable difference between the means of the ETC measure. This is a clear indicator that the periodic and chaotic nature of the tonic and irregular spike trains are characterized by ETC measure while LZ fails to do so. In the cases of Shannon entropy and $SubSym$ measures, mean values of the tonic is higher than that of irregular spiking. This is undesirable (as well as non-intuitive)\footnote{This strange behaviour of Shannon entropy and $SubSym$ measures needs further investigation.} and hence limits the use of these measures for the purpose of characterization.

To further validate the results, the complexity values from each of the windows ($50$ ms duration) were used as sample values and statistical difference was analyzed using confidence interval plots.  Using the same sample values, 2 sample student’s t-test was also done to complement the graphical analysis with the more formal hypothesis test.

\par The t-test results may be summarized as follows:
\begin{itemize}
\item The mean $LZ$ complexity of irregular spiking pattern ($0.62 \pm 0.12$) is not significantly greater ($t_{36}= 0$, \textit{p} $= 0.50$) than that of the tonic spiking pattern ($0.62 \pm 0.10$). 

\item The mean $ETC$ complexity of irregular spiking pattern ($0.28 \pm 0.04$)
 is significantly greater ($t_{36}= 5.84$,  \textit{p} $= 0.00$) than that of the tonic spiking pattern ($0.21 \pm 0.03$).

\item The mean Shannon entropy $H$ of irregular spiking pattern ($0.37 \pm 0.09$) is not significantly greater ($t_{36}= -4.93$,  \textit{p} $= 1.00$) than that of the tonic spiking pattern ($0.47 \pm 0.03$). 

\item The mean $SubSym$ complexity of irregular spiking pattern ($0.65 \pm 0.10$) is not significantly greater ($t_{36} = -6.51$,  \textit{p} $= 1.00$) than that of the tonic spiking pattern ($0.80 \pm 0.01$). 

\end{itemize}

Based on the sample data, at a 5\% significance level (overall error rate) for the statistical test, there is sufficient evidence to conclude that the mean $LZ$, mean Shannon entropy $H$ and mean $SubSym$ complexity measures of irregular spiking patterns are {\it not significantly higher} than those of the corresponding tonic  patterns, while the mean $ETC$ complexity measure of irregular spiking pattern {\it is significantly higher} than that of the tonic pattern. Thus we assert that ETC complexity measure is able to characteristically (statistically) differentiate between the two neuron firing patterns, while the other three complexity measures fail to do so.

\section{Conclusions}
\label{concl}

Complexity has many facets and perspectives, and we have chosen three specific viewpoints - {\it effort-to-describe} (Shannon Entropy and Lempel-Ziv Complexity), {\it effort-to-compress} (ETC) and {\it degree-of-order} (Subsymmetries). Each of these perspectives is distinct and unique, yet there is also overlap amongst them. For example, entropy and lossless compression are deeply related (through Shannon's theorems~\cite{Shannon1948,CoverThomasBook}). Compressible patterns are typically symmetrical whereas random incompressible patterns are asymmetrical. `Subsymmetries' relates to {\it intrinsic} order of arrangements of the symbols that make up a string. It relates to the perspective of visual complexity which is rooted in the cognitive ability of humans to perceive structure/order and distinguish it from the lack of it. A normalized measure based on subsymmetries ($SubSym$) was proposed in this paper. $SubSym$ takes a large number of distinct values for binary strings which is desirable since it allows to distinguish different binary patterns more finely. This new measure deserves to be tested further in various applications. ETC, a recently proposed measure, has intuitive appeal, is easy to compute, and also demonstrates superior performance in discriminating between periodic and chaotic time series. We conclude by noting that {\it complexity} remains a multi-dimensional notion and no single measure can capture it in its entirety. Several perspectives and measures need to work in unison to help us arrive at a better understanding of {\it complexity}.






\medskip
\section*{Acknowledgments}
The authors would like to thank Aditi Kathpalia, PhD scholar (NIAS) for rendering help with simulations pertaining to binary sequences. 

\bibliographystyle{apsrev4-1}
\bibliography{complexsys}

\end{document}